\documentclass[aps,twocolumn,groupedaddress,showpacs,nofootinbib]{revtex4}

\usepackage{graphicx}
\usepackage{hyperref}
\usepackage{dcolumn}
\usepackage{bm}
\usepackage{epsfig,amsbsy,bm,amssymb,amsmath}
\usepackage{xcolor} 
\usepackage{mathtools}

\newcommand{\hs}{\hspace*}
\newcommand{\vs}{\vspace*}
\newcommand{\np}{\newpage}

\newcommand{\eref}[1] {(\ref{#1})}
\newcommand{\Eref}[1] {Eq.~(\ref{#1})}
\newcommand{\Fref}[1] {Fig. \ref{#1}}
\newcommand{\Sref}[1] {Sec.~\ref{#1}}

\newcommand{\nn}{\nonumber}
\newcommand{\be}{\begin{equation}}
\newcommand{\ee}{\end{equation}}
\newcommand{\br}{\begin{eqnarray*}}
\newcommand{\er}{\end{eqnarray*}}
\newcommand{\ba}{\begin{eqnarray}}
\newcommand{\ea}{\end{eqnarray}}
\newcommand{\bp}{\begin{minipage}}
\newcommand{\ep}{\end{minipage}}
\newcommand{\bt}{\begin{tabular}}
\newcommand{\et}{\end{tabular}}

\newcommand{\ms}{\vspace*{-5mm}}
\newcommand{\mms}{\vspace*{-2.5mm}}
\renewcommand{\l}{\lambda}

  \newcommand{\q}{{\bm q}}
  \newcommand{\e}{{\bm e}}
  
 \newcommand{\ig}[1]{\includegraphics[width={#1}]}

\newcommand{\etal}{{\em et al}~}

\renewcommand{\t}{\tau}
\renewcommand{\l}{\lambda}
\renewcommand{\b}{\beta}
\newcommand{\g}{\gamma}
\newcommand{\s}{\sigma}
\renewcommand{\e}{\epsilon}
  
\renewcommand{\q}{\theta}

\newcommand{\three}[6]
{\left(\!\!\!\begin{array}{rrr}
{#1\!}&{#2\!}&{#3\!}\\
{#4\!}&{#5\!}&{#6\!}\\
\end{array}\!\right)}

\newcommand{\threej}[6]{ \begin{pmatrix}			
   #1 & #2 & #3 \\
   #4 & #5 & #6 
\end{pmatrix}}

\newcommand{\dee}{\mathrm{d}}					
\newcommand{\rme}{\mathrm{e}}					
\newcommand{\rmi}{\mathrm{i}}					
\newcommand{\dip}{\hat{{\mathsf{d}}}}			
\newcommand{\rhat}{\hat{{\mathsf{r}}}}			

\DeclarePairedDelimiter\abs{\lvert}{\rvert}		
\DeclarePairedDelimiterX\braket[3]{\langle}{\rangle}{\mathopen{}#1 \delimsize\vert \mathopen{}#2 \delimsize\vert \mathopen{}#3}	
\DeclarePairedDelimiterX\ibraket[2]{\langle}{\rangle}{\mathopen{}#1 \delimsize\vert \mathopen{}#2}	
\DeclarePairedDelimiterX\rbraket[3]{\langle}{\rangle}{\mathopen{}#1 \delimsize\vert\delimsize\vert \mathopen{}#2 \delimsize\vert\delimsize\vert \mathopen{}#3}		
		
\makeatletter
\let\oldabs\abs
\def\abs{\@ifstar{\oldabs}{\oldabs*}}
\makeatother

\begin{document}
\bibliographystyle{apsrev}

\title{ Attosecond Time Delays at Cooper Minima in Valence-Shell
  Photoionization of \\ Alkali and Alkaline-Earth Metal Atoms}

\author{Adam Singor$^{1,2}$, Dmitry V. Fursa$^2$, Igor Bray$^2$}
\author{Anatoli~S. Kheifets$^3*$}

\affiliation{$^{1}$Los Alamos National Laboratory, Los Alamos, NM
  87545, United States}

\affiliation{$^{2}$Department of Physics and Astronomy, Curtin
  University, Perth 6102, Australia}

\affiliation{$^{3}$Research School of Physics, The Australian National
  University, Canberra ACT 2601, Australia}
\email{A.Kheifets@anu.edu.au}

 \date{\today}


\begin{abstract}
Ji \etal [New J. Phys. {\bf 26}, 093014 (2024)] established a direct
link between the photoionization cross section and the attosecond time
delay near Cooper minima (CM) in the valence shells of noble-gas
atoms. This link is based on the analytic properties of the ionization
amplitude in the complex plane of the photoelectron energy, and is
particularly sensitive to the winding number of the amplitude around
the origin of the complex energy plane.
Here, we demonstrate an analogous relation for photoionization of the
valence $ns$ shells of alkali-metal atoms (AMA), from Na ($n=3$) to Cs
($n=6$), as well as alkaline-earth-metal atoms (AEMA), from Mg ($n=3$)
to Ba ($n=6$). To this end, we employ a fully relativistic formalism
that separates the two complementary $ns_{1/2} \to Ep_{1/2}$ and
$Ep_{3/2}$ ionization channels. Each of these channels exhibits a
phase variation close to $\pi$, but in opposite directions, near their
respective Cooper minima. This phase variation vanishes in a
nonrelativistic formulation, where the two channels become degenerate.
For AMA, due to the threshold proximity of the CM, the universal
Coulomb contribution to the time delay must be subtracted. The
remaining component of the time delay is target-specific,
angular-dependent, and accessible through comparative measurements.
\end{abstract}

\maketitle
\section{Introduction}

Prominent minima in atomic photo\-ionization cross sections close to
the threshold were discovered and explained in the mid-twentieth
century \cite{Ditchburn1943,Bates1946,Seaton1951}. Commonly, these
minima are attributed to John Cooper who established the systematics
of this phenomenon \cite{cooper62a,Cooper1964}. An abnormally small
cross section near a Cooper minimum (CM) is due to a sign change of
the radial integral containing the bound and continuous atomic
orbitals.  This
phenomenon is observed in many atomic systems (see Section 4.5 of
\cite{RevModPhys.40.441} and Section 12 of \cite{Starace1983atomic}
for a comprehensive review).

In
atto\-second science, a rapid variation of the phase of the ionization
amplitude near CM is associated with a very large attosecond time
delay which can be detected experimentally
\cite{PhysRevLett.106.143002,0953-4075-47-24-245602,Alexandridi2021,li2025photoionization}. 
In the context of strong-field laser physics, Cooper minima shape high-order
harmonic generation in noble gas atoms (NGA)
\cite{HiguetPRA2011,PhysRevLett.112.153001,Boltaev2020}.

Just by itself, the sign change of the radial integral does not
introduce a phase variation of the photoionization matrix element
except for a sudden jump by $\pi$. For a non-$s$ initial state, it
is due to the coupling of the two ionization channels $n\ell \to E\ell
\pm1$ with their associated scattering phases $\delta_\ell$ that the
net phase of the ionization amplitude varies strongly near the
CM. Such a behavior produces a large energy derivative of the phase
known as the Eisenbud--Wigner--Smith time delay
\cite{Kheifets2023review}.  We will refer to this time delay as the
Wigner, or attosecond, time delay in the following.  A large Wigner
time delay associated with the CM is observed in valence shells of
NGA. In a sub\-valent $ns$ shell, the CM is induced by inter-shell
correlation with the outer $np$ shell. The correlation induced CM is
also associated with a strong phase variation and a large atto\-second
time delay (see a review article \cite{Kheifets2023review} and
references therein). Standing apart is the case of CM in valence $ns$
shells of alkali metal atoms (AMA) from Na ($n=3$) to Cs ($n=6$) and
the alkaline-earth metal atoms (AEMA) from Mg ($n=3$) to Ba
($n=6$). Here the ionization amplitude in the only open
non-relativistic channel $ns\to Ep$ changes its sign momentarily while
the elastic scattering phase $\delta_{\ell=1}$ remains smooth. Hence
no sizable atto\-second time delay can be associated with the
corresponding CM by a simple energy differentiation of the phase.

It was realized rather early that the relativistic formulation changes
the CM description in AMA rather profoundly
\cite{Ong1978,Fink1986}.  The spin-orbit splitting is
responsible for a nonzero CM 
%
%
because the two spin-orbit split ionization channels $ns_{1/2}\to
Ep_{1/2}$ and $Ep_{3/2}$ pass through their respective CM at a
slightly different energy. This also causes the angular dependence of
the photo\-ionization cross section to become non-trivial with the
angular asymmetry parameter $\b_{ns_{1/2}}\ne 2$. A similar effect is
present in sub\-valent $ns$ shells of NGA \cite{Decleva2020}.  Each of
the two competing relativistic channels displays a phase variation by
one unit of $\pi$ near their respective Cooper minimum. This variation
is associated with a peak in the corresponding time delay which is
slightly offset between the two channels. This behavior results in
angular dependence of the net time delay.

Such  behavior is also present in the case of AMA and AEMA which we
analyze in the present work. Our analysis reveals a fundamental
difference between the metal and noble gas atoms. In noble gases, the
phase variation of the ionization amplitude in the $ns_{1/2}\to
Ep_{1/2}$ and $Ep_{3/2}$ channels is slightly offset in energy near
their respective CM but occurs in the {\em same} direction. Thus in a
non-relativistic formulation with just one $ns\to Ep$ ionization
channel the phase variation remains sizable. On the contrary, in the
metal atoms, the phases of the $ns_{1/2}\to Ep_{1/2}$ and $Ep_{3/2}$
amplitudes vary in the {\em opposite} directions. The phase jump
vanishes in the non-relativistic formulation and the CM associated
time delay remains negligible on the background of the smooth Coulomb
contribution.

After establishing the phase behavior in the two relativistic channels
we relate it with the cross section variation near the CM by way of
the logarithmic Hilbert transform (LHT) \cite{Ji2024}. More
specifically, the cross section is fitted with the Fano-like ansatz
and thus deduced Fano parameters are converted analytically to the
phase and time delay. The Fano-LHT formalism allows for the two
alternative phase and time delay values depending on the winding
number of the ionization amplitude in the complex photo\-electron energy
plane. We demonstrate that the $j=1/2$ and $3/2$ ionization channels
correspond to the two different winding numbers that leads to their phase
variation occurring in the opposite directions. 

Our numerical simulations are based on the two complementary
approaches. We treat the closed-shell AEMA with the relativistic
random phase approximation (RRPA)
\cite{PhysRevA.20.964,1402-4896-21-3-4-029}. In principle, RRPA can
also be applied to open-shell AMA as was demonstrated by
\citet{Fink1986}. We attempt to repeat these calculations but treat
our results with caution as no clear numerical recipes were given in
\cite{Fink1986} and our adaptation of the RRPA code
\cite{PhysRevA.20.964} was taken at our own responsibility. A more
common approach to the open-shell systems like AMA is taken by
the model-potential method (MPM) \cite{Weisheit1972,Norcross1973}
which accounts for core polarization by empirical parameterization as
well as for spin-orbit interaction.  Here we follow a similar
route. Details of our MPM are presented in earlier works
\cite{Singor2021,Singor2022}. In brief, the method starts from a
frozen core Dirac-Fock (DF) basis. Many-electron correlation is
introduced in the dipole polarization potential and the electron
dipole operator. Thus obtained photo\-ionization cross sections and
angular asymmetry $\b$ parameters are compared favorably with the
known literature values. This allows us to proceed with confidence and
to evaluate the energy and angle dependence of the time delay in the
studied sequence of AMA.
%

The rest of the paper is organized as follows. In \Sref{CS} we present
relativistic expressions for the photo\-ionization cross section $\s$,
the angular asymmetry $\b$ parameter and the atto\-second time delay
$\t$. The logarithmic Hilbert transform technique (LHT) is described
in \Sref{LHT}. Then we proceed with outlining our numerical
techniques: RRPA in \Sref{RRPA} and MPM in \Sref{MPM}. Our numerical
results are presented in \Sref{Results}.  In \Sref{RRPAres} we present
our RRPA calculations and in \Sref{MPMres} we outline our MPM results.
Finally, we summarize in \Sref{Summary} by drawing main
conclusions of the present study.

\section{General formalism}
\label{Formalism}

\subsection{Cross section, angular asymmetry and time delay}
\label{CS}

We follow the derivation of our previous works
\cite{PhysRevA.94.013423,Decleva2020} where relativistic formulation
of the angular dependent time delay was established. 
%
%
This derivation is based on the decomposition of the ionization
amplitude, in terms of the total angular-momentum projection $m$ and
the spin projection $\nu$, for the electric ($\l=1$) dipole ($J=1$)
transition with linearly polarized radiation ($M=0$).
\ba 
T_{J=1,M=0}^{(\lambda=1)}
\equiv
{[T_{1 0}^{1\nu}]}_{m}=
\sum_{\bar{\kappa}\bar{m}}
C^{jm}_{\ell,m-\nu,1/2,\nu}
Y_{\ell m-\nu} (\hat{k})
\chi_{\nu} 
\nn\\&&\hs{-7.5cm}\times
(-1)^{\bar j+ j +J + \bar j-\bar m}
\left(
\begin{array}{c c c} \bar{j} & J & j \\ -\bar{m} & M & m \end{array}
\right)
 D_{\ell j \to \bar{\ell}\bar{j}}
\ .
\label{DLJ}
\ea
Here the reduced dipole matrix element $D_{\ell j \to
  \bar{\ell}\bar{j}}$ is fully stripped of the angular dependence and
the spin is described by a two-component spinor $\chi_{\nu}$.  The
spherical harmonic $Y_{\ell m}(\hat k)$ is evaluated in the direction
of the photoelectron emission with the quantization direction $\hat z$
chosen along the polarization axis. While $\chi_{\nu}$ and
$Y_{\ell m}$ are known analytically,  $D_{\ell j \to \bar{\ell}\bar{j}}$ will
be evaluated numerically in the following.

In the present case of the AEMA and AMA with an $ns_{1/2}$ initial state,
\Eref{DLJ} leads to the two following expressions:
\ba
\label{amplitudes}
&&\\\nn
{[T_{1 0}^{1+}]}_{ns_{1/2}}^{m=\frac12}
&=& 
{-1\over3\sqrt2} Y_{10}D_{ Ep_{1/2}}
- {1\over3} Y_{10} D_{ Ep_{3/2}} \ ,
\\
{[T_{1 0}^{1-}]}_{ns_{1/2}}^{m=\frac12}
&=& \nn
{1\over3} Y_{11}D_{ Ep_{1/2}}
- {1\over3\sqrt2} Y_{11} D_{ Ep_{3/2}}
\ .
\nn
\ea
Here, for brevity of notation, we introduce the shortcuts $[T_{1
    0}^{1\pm}]\equiv [T_{1 0}^{1\nu=\pm\frac12}]$, $D_{Ep_j}\equiv
D_{ns_{1/2}\to Ep_{j}}$ and drop the argument of the spherical
harmonics.
The analogous expressions with  $m=-1/2$  will have a
similar structure with the simultaneous inversion of the spin
projection $T^+\leftrightarrow T^-$ and the second index of the
spherical harmonic $Y_{21}\to Y_{2-1}$.

Each amplitude in \Eref{amplitudes} has its own associated
photoelectron group delay (Wigner time delay) defined as
\be
\t^\pm =  \frac{d\eta^\pm}{dE}
\ \ , \ \ 
\eta^\pm = \tan^{-1}\left[
{{\rm Im}\, T_{1 0}^{1\pm} \over
{\rm Re}\, T_{1 0}^{1\pm}}
\right]
\ .
\ee
The spin-averaged time delay can be expressed as a weighted sum
\be
\bar\tau_{ns_{1/2}} = {
\tau^{m=\frac12,+}_{ns_{1/2}} 
\left|[T^{1+}_{10}]_{ns_{1/2}}^{m=\frac12}\right|^2
+
\left|[T^{1-}_{10}]_{ns_{1/2}}^{m=\frac12}\right|^2
\tau^{m=\frac12,-}_{ns_{1/2}} 
\over
\left|[T^{1+}_{10}]_{ns_{1/2}}^{m=\frac12}\right|^2
+
\left|[T^{1-}_{10}]_{ns_{1/2}}^{m=\frac12}\right|^2
}
\ .
\ee
Knowing the reduced dipole matrix elements we can also evaluate the angular
asymmetry $\beta$ parameter expressed as \cite{A90}
\be
\label{beta}
\beta_{ns_{1/2}} = 
{
|D_{Ep_{3/2}}|^2
+2\sqrt2 {\rm Re}[D^*_{Ep_{3/2}}D_{Ep_{1/2}}]
\over 
|D_{Ep_{3/2}}|^2 +
|D_{Ep_{1/2}}|^2
}
\ . 
\ee 
Here we use a shortcut $D_{Ep_j}\equiv D_{ns_{1/2}\to Ep_{j}}$ where
the right-hand side is defined by \Eref{DLJ}.
We note that $2\ge\b_{ns_{1/2}}\ge -1$ attains the upper bound at
$R={D_{Ep_{3/2}}/ D_{Ep_{1/2}}}=\sqrt2$ and the lower bound at
$R=-1/\sqrt2$. While $R=\sqrt2$ corresponds to a weakly-relativistic
limit of two degenerate $Ep_{3/2}$ and $Ep_{1/2}$ channels,
$R=-1/\sqrt2$ can only be attained near the CM of the normally stronger
$Ep_{3/2}$ channel where the sign of its dipole matrix element is
flipped. 

The total photoionization cross section which has the following
form in the length gauge of the electromagnetic interaction
\cite{PhysRevA.20.964}:
\be
\label{sigma}
\sigma_{ns_{1/2}} = {4\pi^2\over 3}\alpha a_0^2\omega 
\Big[
|D_{Ep_{3/2}}|^2 +
|D_{Ep_{1/2}}|^2
\Big]
\ .
\ee
Here $\alpha$ is the fine structure constant, $a_0$ is the Bohr radius
and $\omega$ is the photon frequency. The atomic units with
$e=m=\hbar=1$ are in use here and throughout.

\subsection{Ionization amplitudes and phases}
\label{LHT}
\mms

As was shown by \citet{Ji2024}, the logarithmic Hilbert transform
(LHT) links the photo\-ionization cross section near the Cooper
minimum with the phase of the corresponding ionization amplitude and
the associated time delay. The proviso of this relation is that the
cross section is the square of the unique ionization amplitude rather
than an incoherent sum of squared amplitudes over several ionization
channels. Therefore the cross section-to-phase conversion cannot be
applied directly to \Eref{sigma} in which the sum over $j=1/2$ and
$3/2$ is conducted. Instead, we apply it separately to the spin-split cross sections expressed via the corresponding amplitudes
$
\sigma_j\propto |D_{Ep_j}|^2
\ .
$
These cross sections, along with the total cross section $\sigma=
\sigma_{1/2}+\sigma_{3/2}$ are exhibited in \Fref{Fig2}. Each of the 
$\sigma_j$ cross sections are fitted with the Fano
ansatz~\cite{PhysRev.137.A1364}:
\be 
\label{Fano}
\sigma_{\mathrm{F}}(\e) = \sigma_{0} \left[ \rho^2
\frac{(q+\e)^2}{1+\e^2} + 1 -\rho^2\right] 
\ \ , \ \ 
\e= {E-E_0\over \Gamma/2} \ .  
\ee
In \Eref{Fano}, $\sigma_0$ describes the flat background, $\e$ is a
detuning from the resonance center $E_0$ measured in units of the
resonance half width $\Gamma$ and $q$ is the Fano shape index.  An
additional correlation factor $\rho$ is required when several
continuum channels are degenerate at the resonance energy \cite{PhysRev.137.A1364}.

An alternative Fano parametrization suggested by \citet{Ji2024}
recombines the two factors $q$ and $\rho$ into the complex shape index
$Q+\rmi \gamma$ resulting in the following cross section parameterization:
\be 
\sigma_{\mathrm{F}}(\e) = \sigma_0
\frac{(\e+Q)^2+\gamma^2}{\e^2+1}.  
    \label{eq:sigma_general}
\ee
An analytic amplitude $\mathfrak{D}(\e)$ can be defined such that
$\abs{\mathfrak{D}(\e)}^2 = \sigma_{\mathrm{F}}(\e)$, and thus it can be
written as
\be
\label{c-amplitude}
\mathfrak{D}(\e) = \sqrt{\sigma_0} \,\frac{\e+Q
  +\rmi\gamma}{\e +\rmi} 
\ . 
\ee
The cross section expression \eref{eq:sigma_general} leaves the sign
of $\gamma=\pm|\gamma|$ undetermined. This expression also allows for
an arbitrary phase shift of the amplitude \eref{c-amplitude}.  The
sign of $\g$ is crucial for the phase variation of the amplitude
\eref{c-amplitude} near the CM. The positive sign $\gamma>0$
corresponds to the positive $\pi$ variation of the phase when the
photo\-electron energy increases. This results in the positive Wigner
time delay near the CM. Conversely, the negative sign $\gamma<0$ means
the phase making a $\pi$ downturn and the time delay becoming
negative. Such a profound difference can be explained by the analytic
properties of the amplitudes \eref{c-amplitude} in the complex
photo\-electron energy plane \cite{Ji2024,Kheifets2023review}.  The
positive sign $\g>0$ corresponds to the amplitude encircling the
origin when $\e\to0$ thus acquiring an additional node and making the
logarithmic derivative singular. Such a singularity needs to be
isolated and integrated out when performing the LHT. This results in
the sign reversal of the phase and time delay. In the opposite case of
$\g<0$, the amplitude misses the origin near the CM and the
logarithmic derivative remains regular. So the LHT can be applied
straightforwardly. These two alternative cases are differentiated by
the winding numbers of 1 and 0 corresponding to the number of turns
made by the amplitude in the complex energy plane near the CM.

While the sign of $\g$ and the resulting phase variation is ambiguous
in the analytic amplitudes, there is no such ambiguity in the
numerical amplitudes. The latter amplitudes, however, are plagued by
the Coulomb phase which becomes divergent near the
threshold. Fortunately, the Coulomb phase is known analytically and
can be subtracted out from the dipole matrix elements leaving only the
distorting phase due to the short-ranged non-Coulomb part of the
potential. After this subtraction, the winding numbers of the
numerical and analytic amplitudes can be matched.

\np
\section{Numerical techniques}
\label{Numerical}

\mms
\subsection{Relativistic Random Phase Approximation}
\label{RRPA}

Relativistic random-phase approximation (RRPA) \cite{PhysRevA.20.964},
as well as its nonrelativistic counterpart\footnote{The
  acronym RPAE stands for random phase approximation with
  exchange. The exchange is also fully taken into account in the RRPA
  but not highlighted in its acronym.}  (RPAE) \cite{A90}, serve to
isolate the dominant mechanism by which a nondegenerate closed-shell
atomic system responds to a weak electromagnetic field. This response
manifests itself in the creation of multiple electron-hole pairs.  The
ground state (electron-hole vacuum) is described in the Dirac-Fock
(DF) or the Hartree-Fock (HF) approximations in the relativistic and
non-relativistic cases, respectively. Both the DF and HF
approximations make the ground state variationally stable to creation
of the electron-hole pair and it is the external electromagnetic field
that is necessary to initiate this process. The RRPA and RPAE are
essentially multi-channel theories which take into account the process
of inter-shell correlation.

RRPA is derived by linearizing the time-dependent DF equations in
powers of the external field \cite{PhysRevA.20.964}. In the DF basis,
the amplitude for a transition from the ground state ($u_{i}$) to an
excited state ($\omega_{i\pm}$), induced by a time varying external
field $v_+\rme^{-\rmi\omega t} + v_-\rme^{\rmi\omega t}$ is given by
\be 
T=\sum_{i=1}^{N}  \int \dee^{3}r\,(\omega_{i+}^{\dagger}\vec{\alpha}\cdot\vec{A}
u_{i}+u_{i}^{\dagger} \vec{\alpha}\cdot\vec{A}\omega_{i-})
\ .  
\ee
Here the electromagnetic interaction is written in the Coulomb gauge
and expressed in terms of the Pauli spin matrices 
$\vec{\alpha} =\left(
\begin{array}{c c} 0 & \vec{\sigma}\\ \vec{\sigma} & 0 \end{array}
\right)$
and the vector potential $\vec{A}$.

In a single active electron approximation, the multipole transition
amplitude is reduced to 
\be 
T_{JM}^{(\lambda)}=\int
\dee^{3}r\,{\omega_{i+}^{\dagger}\vec{\alpha}\cdot\vec{a}_{JM}^{\lambda}u_{i}}
\ ,
\ee
where the indices $J$ and $M$ are the photon angular momentum and its
projection and $\lambda = 1$ or 0 for electric or magnetic multipoles,
respectively.  With the radial and angular variables separation
expressed by \Eref{DLJ}, the transition from an initial state
characterized by the quantum numbers $\ell j$ to a final continuum state
$\bar{\ell}\bar{j}$ is described by  the reduced matrix
element modified by the phase factors:
\be
D_{\ell j \to \bar{\ell}\bar{j}}= \rmi^{\ell-\bar{\ell}}
\rme^{i\delta_{\bar{\kappa}}}
\left\langle \bar{a} \|Q_J^{(\lambda)} \|a\right\rangle \ .
\label{phase_factor}
\ee
This reduced matrix element between the initial
state $a = (n\kappa)$ and a final energy-scale normalized state $\bar{a}= (E,
\bar\kappa)$ is written as
\ba
\hs{-1cm}
\left\langle \bar{a} \|Q_J^{(\lambda)}\|a\right\rangle
&=&
(-1)^{j+1/2}[\bar j][j]
\three{j}{\bar j} {J}
{-1/2} {1/2}{0}
\nn\\&&\times \
\pi(\bar \ell,\ell,J-\l+1)
R^{(\l)}_J(\bar a,a)
\ .
\label{reduced}
\ea
Here $\pi(\bar \ell,\ell,J-\l+1)=1$ or 0 for $\bar \ell+\ell+J-\l+1$
even or odd, respectively, $[j]=\sqrt{2j+1}$, and $R^{(\l)}_J(\bar
a,a)$ is the radial integral.

\mms
\subsection{Model potential method}
\label{MPM}

We model an AMA as a single active electron system in a local central
potential produced by the frozen electron core.  The core orbitals are
obtained by the self-consistent field DF
method~\cite{dyall1989}.
The local central potential is represented by the three terms
\cite{bartschat1996COMP}:
\begin{equation}
V^{(\kappa)}(r) = V_{\mathrm{d}}(r) + V^{(\kappa)}_{\mathrm{e}}(r) +
V^{(\kappa)}_{\mathrm{p}}(r)
\ . 
\label{eq:model_potential}
\end{equation}
The direct term $V_{\mathrm{d}}$ accounts for the
Coulomb interaction of the valence electron with the nuclei and
the frozen-core electrons:
\begin{equation}
V_{\mathrm{d}}(r) = -\frac{Z}{r} + \sum_{n_{\mathrm{c}}
  \kappa_{\mathrm{c}}} (2j_{\mathrm{c}}+1) \int_0^\infty  \frac{\abs{
    \phi_{n_{\mathrm{c}} \kappa_{\mathrm{c}}}(r')
  }^{2}}{\mathrm{max}(r,r')}\, \dee r'
\ . 
\label{eq:Vdir}
\end{equation}
Here $Z$ is the charge of the nucleus and $\phi_{n_\mathrm{c}
  \kappa_{\mathrm{c}}}(r)$ are the reduced radial wave functions for
core orbitals.  
The localized exchange term $ V^{(\kappa)}_{\mathrm{e}}$ is
approximated  by the form suggested by \citet{furness1973}:
\begin{align}
V^{(\kappa)}_{\mathrm{e}}(r) &= - \frac{\alpha_{\mathrm{exch}}^{(\kappa)}}{2} \bigg[ \sqrt{ V_{\mathrm{d}}(r)^2 + 4 \pi \rho(r)}    +V_{\mathrm{d}}(r) \bigg], \label{eq:locex}
\end{align}
where 
\begin{equation}
\rho(r) = \sum_{n_{\mathrm{c}} \kappa_{\mathrm{c}}} (2j_{\mathrm{c}}+1) \frac{\abs{\phi_{n_{\mathrm{c}} \kappa_{\mathrm{c}}}(r)}^{2}}{4\pi r^2}
\end{equation}
is the electron density of the core and
$\alpha_{\mathrm{exch}}^{(\kappa)}$ is a parameter that is chosen, for
each relativistic quantum number $\kappa$, to ensure the local
exchange potential is equivalent to its non-local counterpart.
Effect of the valence electron on the core is accounted for 
by a dipole polarization potential~\cite{norcross1974},
which is given by
\begin{equation}
V^{(\kappa)}_{\mathrm{p}}(r) = - \frac{\alpha_{\mathrm{D}}}{2r^4}
\left\{ 1 - \exp \!\left[- \!\left(
  \frac{r}{r_{\mathrm{c}}^{(\kappa)}} \right)^6 \right]
\right\}
\ . 
\label{eq:Vdippol}
\end{equation}
Here $\alpha_{\mathrm{D}}$ is the static dipole polarizability of the
core ion, and $r_{\mathrm{c}}^{(\kappa)}$ is a cut-off radius that is
chosen for each $\kappa$ by fitting to experimental energy
levels~\cite{nistASD}.

The electric dipole operator $\dip$ must also be modified to account
for the polarization of the core by the valence electron.  To do this
we have performed calculations using the modified length form of the
dipole operator~\cite{hameed1968,caves1972,mitroy1988}:
\begin{equation}
\dip^{\mathrm{mod}}= \rhat \left\{ 1 - \frac{\alpha_{\mathrm{D}}}{r^3} \sqrt{ 1 - \exp\!\left[ - \!\left( \frac{r}{\tilde{r}_{\mathrm{c}}} \right)^6\right]}\,\right\},
\label{eq:modDip}
\end{equation}
where $\alpha_{\mathrm{D}}$ is the static dipole polarizability of the
core ion, and the cut-off radius $\tilde{r}_{\mathrm{c}}$ is chosen to
reproduce the weighted average of the experimental resonant transition
oscillator strengths~\cite{nistASD}.  Values for the free parameters
in our local central potential and modified dipole operator are
tabulated in Ref.~\cite{Singor2021}.

Continuum states of the valence electron are obtained by solving a
pair of coupled first-order differential equations for the large and
small components of the radial wave function. These
differential equations were solved using a 5-point Adams--Moulton
predictor-corrector method~\cite{sienkiewicz1987}, which has an error
$\mathcal{O}(h^{7})$ where $h$ is the step size between two
consecutive grid points.

We employ the semi-relativistic Pauli approximation to evaluate matrix
elements of the modified dipole operator, \Eref{eq:modDip},
\begin{align}
\braket{\bar n\bar\kappa\bar m}{\dip^{\mathrm{mod}}_{q}}{n\kappa m} &=
(-1)^{\bar j-\bar m}\threej{\bar j}{1}{j}{-\bar m}{q}{m} 
\ .
D_{n\kappa\to \bar n \bar \kappa} \ .
\end{align}
The reduced dipole matrix elements $D_{n\kappa\to \bar n \bar \kappa}$
are substituted into the ionization amplitudes \Eref{amplitudes} to
obtain the $\s$, $\b$ and $\t$ parameters.

\ms
\section{Numerical results}
\label{Results}
\mms
\subsection{RRPA calculations}
\label{RRPAres}
\subsubsection{Alkaline-earth metal atoms}

The multi-channel RRPA calculations for AEMA comprise 7 relativistic
channels originated from the valence $ns$ and subvalent $(n-1)p$
atomic shells. The most prominent are the $ns_{1/2}\to Ep_{1/2}$ and
$Ep_{3/2}$ channels. Transitions from the sub\-valent shell affects
ionization of the valence shell insignificantly at the photon energy
close to its threshold. This corroborates the use of single-channel
methods such as MPM. 

\begin{figure*}
\hs{-1cm}
\ig{1.1\textwidth}{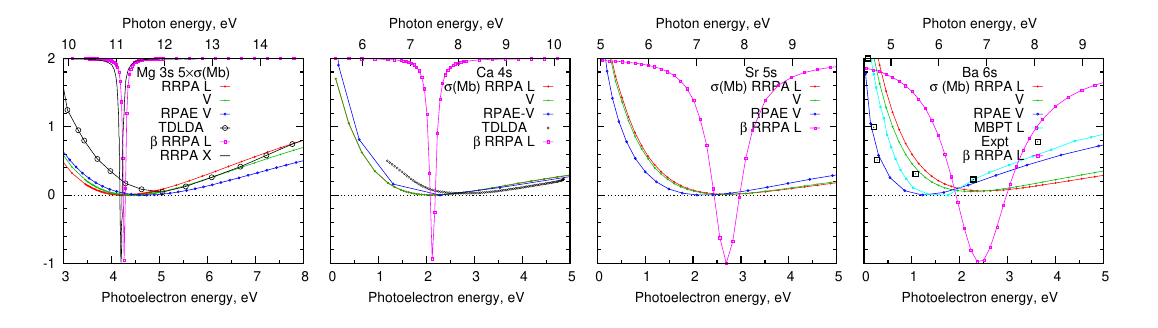}


\caption{The photoionization cross section $\s$ (in Mb) and the
  angular asymmetry $\b$ parameter (dimensionless) in Mg $3s$, Ca $4s$, Sr $5s$ and
  Ba $6s$ are displayed in several panels from left to right. The
  present RRPA calculations in the length (L) and velocity (V) gauges
  are displayed for the cross section and the L gauge is used for the
  $\b$ parameter. The non-relativistic RPAE calculations for the
  cross section are shown in the V gauge. The time-dependent local
  density approximation (TDLDA) is used to evaluate the cross sections
  in Mg $3s$ and Ca $4s$ \cite{Stener1997}. RRPA X results (gauge not
  reported) for the $\b$ parameter in Mg  \cite{Pradhan2011}
  are also shown. In the case of Ba $6s$, we display the many-body
  perturbation theory (MBPT) result for the cross section
  \cite{Kutzner1990} along with the few experimental points reported
  by \citet{Hudson1970}.
\label{Fig1}}
\ms
\end{figure*}

The photoionization cross sections $\s$ (in Mb, scaled on the left
panel) and angular asymmetry $\b$ parameters (dimensionless) in Mg
$3s$, Ca $4s$, Sr $5s$ and Ba $6s$ are displayed in \Fref{Fig1} (from
left to right)
The present RRPA calculations in the length (L) and velocity (V)
gauges are displayed for the cross section; the L gauge is used for
the $\b$ parameter. The gauge convergence is good in RRPA and the $\b$
parameters in the L and V gauges are barely distinguishable.  The
non-relativistic RPAE calculations for the cross section are shown in
the V gauge. They are close to both gauges of the RRPA calculation but
the L gauge in the RPAE (not shown) is further apart.  The angular
asymmetry parameter is exactly $\beta=2$ in the non-relativistic case,
and deviates from this value only when relativistic corrections are
included.  In the relativistic calculations, the asymmetry parameter
drops very sharply to $\b=-1$ near the CM. This drop is considerably
deeper than in NGA \cite{Decleva2020}.  The width of this drop
indicates the spacing between the Cooper minima in the two
relativistic $Ep_{1/2}$ and $Ep_{3/2}$ channels. This spacing is
increasing gradually towards the heavy end of the AEMA sequence where
the relativistic effects become stronger because of a large nuclear
charge. This is depicted more graphically in the top row of
\Fref{Fig2}.
Several literature values are also displayed in \Fref{Fig1} for the
sake of comparison. This comparison indicates a fair consistency of
the present results with the previous reports.

\begin{figure*}
\vs{-5mm}
\hs{-2cm}
\ig{1.2\textwidth}{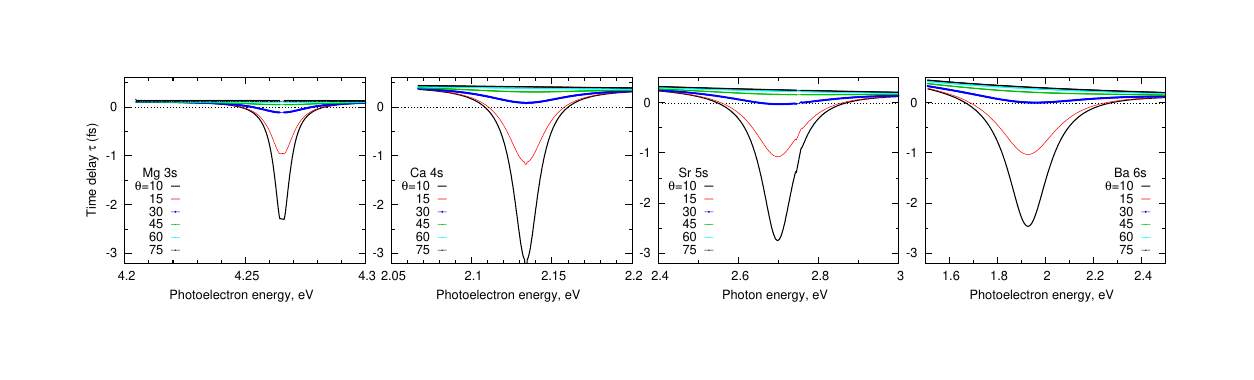}
\vs{-15mm}



\caption{
Photoemission time delay $\t$ (in fs) in Mg $3s$, Ca $4s$, Sr $5s$ and Ba
$6s$ (from left to right) at various emission angles $\q$ (in degrees)
relative to the polarization direction.  Present RRPA calculations in
the length L gauge are displayed. 
\label{Fig3}}

\end{figure*}


The atto\-second time delay across the AEMA series is displayed in
\Fref{Fig3}. Because of the close competition between the $Ep_{1/2}$
and $Ep_{3/2}$ ionization channels, this time delay is very
anisotropic peaking strongly in the polarization direction.  The
broadening of the $\b$ drop in \Fref{Fig1} is mirrored by a similar
broadening in the time delay as a function of the photo\-electron
energy.  The origin of both effects is similar as the relativistic
effects become stronger for the heavier AEMA driving the CM in the
$j=1/2$ and $3/2$ channels wider apart. The Coulomb contribution of the time
delay away from the CM becomes more noticeable as the CM moves closer
towards the threshold. This effect is most visible in Ba $6s$.

The phase and amplitude analysis in the AEMA series is illustrated in
\Fref{Fig2}. The top row of panels exhibits the partial
photo\-ionization cross sections $\s_j$ with $j=1/2$ and $3/2$. Each
of these cross sections is fitted with the Fano ansatz
\Eref{Fano}. This fit returns the two sets of Fano parameters. Aided
by Eqs.~(\ref{eq:sigma_general}--\ref{c-amplitude}), these parameters
are converted to the corresponding complex ionization amplitudes with
the alternative sign of $\g=\pm|\g|$. The phases of these amplitudes
are displayed in the middle row of panels in \Fref{Fig2}. In the same
graphs we show the phases of the corresponding matrix elements
$D_{ns\to Ep_{1/2}}$ $D_{ns\to Ep_{3/2}}$. The $j=1/2$ phase is
shifted downwards by one unit of $\pi$ to fit better to the scale of
the figure. We note that the $j=1/2$ and $3/2$ phases should be
identical in the weakly-relativistic limit. The phases of the analytic
amplitudes \eref{c-amplitude} are shifted by constant values to match
their numerical counterparts. After this shift, we observe the close
proximity of both analytic and numeric amplitudes in the lightest Mg
atom. This proximity starts to deteriorate towards the heavy end of
the AEMA series. In addition, due to the proximity to the ionization
threshold, the Coulomb phase becomes noticeable and tilts the numeric
phases away from the CM.

The bottom row of panels in \Fref{Fig3} exhibits the winding of the
numeric and analytic amplitudes around the origin of the complex plane
of the photo\-electron energy. The scale of the figures is selected to
show how the amplitudes pass near the CM. In addition, the insets on
all the panels, except for the rightmost Ba atom, display the same
amplitudes on the broader photo\-electron energy scale. From this row
of panels, we observe that the winding around the origin is the same
in the $j=1/2$ numeric amplitude and the $+\g$ analytic amplitude,
whereas the same correspondence is seen between the $j=3/2$ numeric
amplitude and its $-\g$ analytic counterpart. The former pair does not
encircle the origin whereas the latter pair does.

\subsubsection{Alkali metal atoms}

As outlined in \Sref{RRPA}, the RRPA method is designed primarily for
closed shell atomic systems and cannot be applied directly to open
shell AMA. Nevertheless, such an application was reported by
\citet{Fink1986}. They did not provide justification for such an
application nor did they give any explicit numeric recipes that we
could follow. We modified the RRPA code by
\citet{PhysRevA.20.964} in two aspects. Firstly, we halved the valence
shell occupation in \Eref{sigma}. Second, we did the same with the
direct potential \Eref{eq:Vdir} driving the RRPA equations. In
addition, the DF equations were solved self-consistently for the core
which was then frozen when calculating the valence shell electron state.

With these modifications we were able to obtain sensible results for
open-shell AMA. These results are illustrated in \Fref{Fig4} where we
display the $\s$, $\b$ and $\t$ parameters in Na $3s$.  We also
illustrate the associated phase-amplitude analysis in the same
figure. The $\s$ and $\b$ parameters from \citet{Fink1986} 
displayed in the left panel are in fair agreement with the present
results. As in AEMA, the gauge convergence is good in RRPA but less so
in RPAE. Nevertheless, the latter cross section in the V gauge is
sufficiently close to both RRPA gauges.


\begin{figure*}

\vs{-5mm}
\hs{-2cm}
\ig{1.2\textwidth}{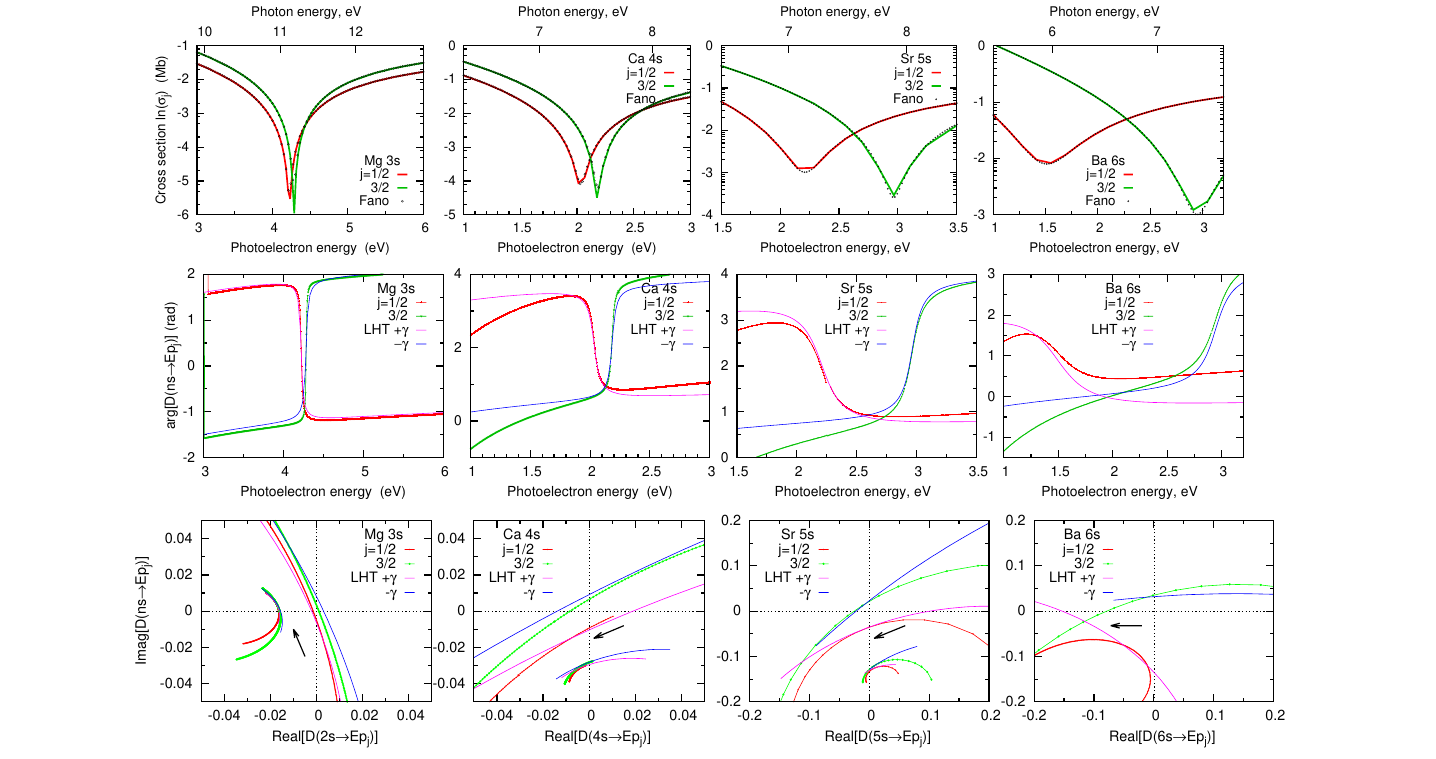}






\caption{The partial photo\-ionization cross sections $\s_j$ (top) and
  phases of the dipole matrix elements $\arg[D_{ns\to Ep_{j}}]$
  (middle) in Mg $3s$, Ca $4s$, Sr $5s$ and Ba $6s$ (from left to
  right). The phases in the $j=1/2$ channel are shifted downwards by
  $\pi$ to make a more graphical comparison with the $j=3/2$
  channel. The phases of the analytic amplitude \eref{c-amplitude}
  with $\g=\pm|\g|$ are also shown. The bottom row illustrates the
  winding of the numeric and analytic amplitudes around the
  origin. The arrows point to the winding direction corresponding to
  the increasing photo\-electron energy. The insets display the
  amplitudes on a broader energy scale.  The present RRPA calculations
  in the length L gauge are displayed.
\label{Fig2}}

\end{figure*}

\begin{figure*}

\vs{-5mm}
\hs{-0.5cm}
\ig{1.05\textwidth}{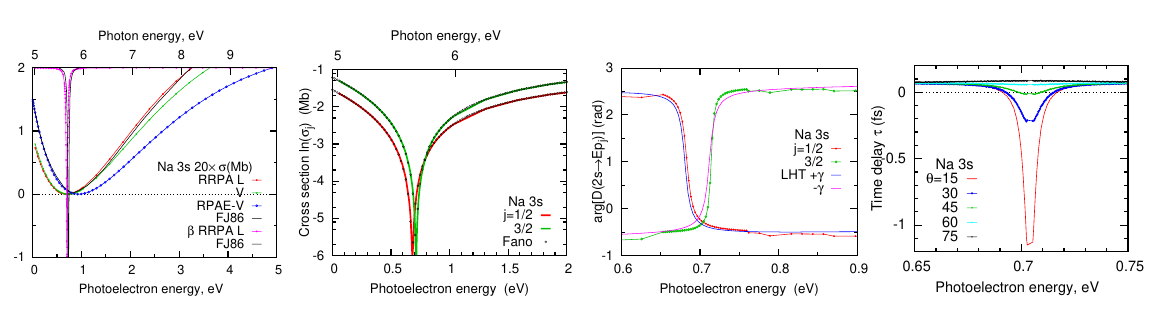}



\caption{The total photo\-ionization cross section and the angular
  asymmetry $\b$ parameter, the partial cross sections $\s_j$, the
  phases in the $j=1/2$ and $3/2$ channels, and the angle-dependent
  time delay in Na $3s$ (from left to right). The $\s$ and $\b$
  parameters from \citet{Fink1986} are shown in the left panel.
\label{Fig4}}

\end{figure*}

\begin{figure*}
\hs{-5mm}\ig{18cm}{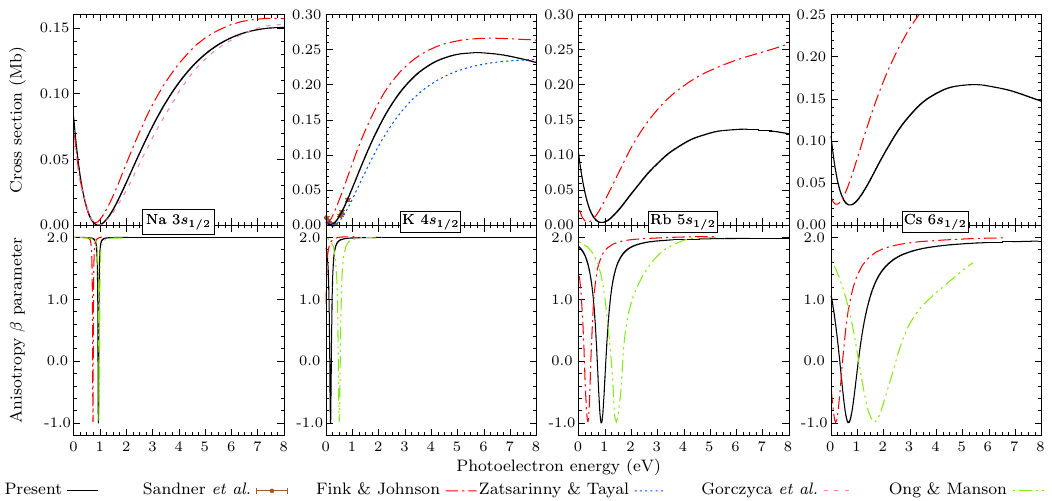}

\caption{ Photoionization cross section (top row) and angular
  asymmetry $\b$ parameter (bottom row) in Na $3s$, K $4s$, Rb $5s$
  and Cs $6s$ (from left to right) as functions of the photo\-electron
  energy. The present calculations are compared with literature
  values by \citet{Ong1978}, \citet{sandner1981},
  \citet{Fink1986}, \citet{zatsarinny2010}  and \citet{Gorczyca2024}.  
\label{Fig5}}
\end{figure*}

\begin{figure*}

\vs{-5mm}
\hs{-10mm}\ig{1.08\textwidth}{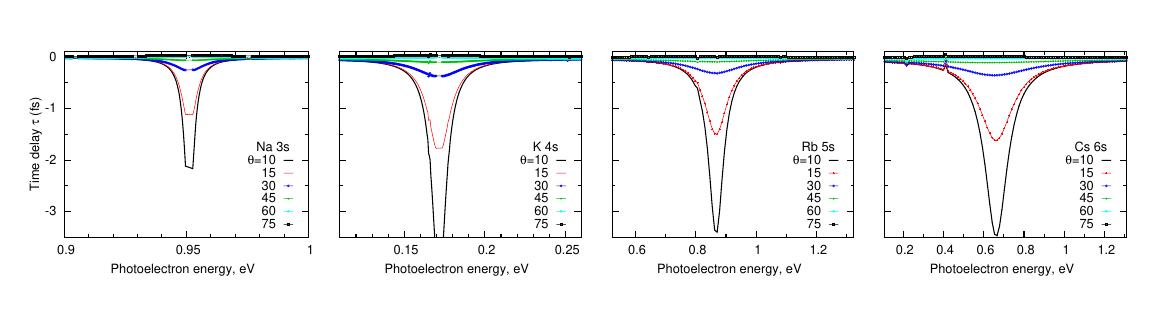}
\vs{-5mm}



\caption{Photoemission time delay $\t$ (in fs) in Na $3s$, K $4s$, Rb
  $5s$ and Cs $6s$ (from left to right) at various emission angles
  $\q$ (in degrees) relative to the polarization direction. The smooth
  Coulomb component of the time delay is subtracted. The present MPM
  results are displayed.
\label{Fig6}}
\end{figure*}


\subsection{MPM calculations}
\label{MPMres}

We conduct the MPM calculations for the AMA series and our results are
shown in \Fref{Fig5}.  The figure displays the photo\-ionization
cross sections (in Mb, the top row) and the angular asymmetry $\b$
parameters (dimensionless, the bottom row) for Na $3s$, K $4s$, Rb
$5s$ and Cs $6s$ (from left to right).  The presently calculated cross
sections are in excellent agreement with the \textit{ab initio}
calculations of \citet{Gorczyca2024} for sodium, particularly in the
region of the Cooper minimum.
Good agreement with the \textit{ab initio} calculations of
\citet{zatsarinny2010} and the measurements of \citet{sandner1981} are
found for potassium.  Overall agreement with the relativistic
random-phase approximation (RRPA) results of \citet{Fink1986}
is poor, particularly with respect to the position of the Cooper
minimum.
There are some old experiments by
\citet{hudson1964,hudson1965,hudson1965a,hudson1967,marr1968};
however, their target samples also contained alkali metal
dimers. Hence, their cross sections contain contributions from these
dimers leading to incorrect magnitudes and additional features in
their cross sections. Nearly all the authors who reported
photo\-ionization calculations for AMA have pointed out the
problems with these experiments and the need for new experiments.
There have been several attempts to modify and rescale the
experimental cross sections to remove the dimer contributions, but
this has not been particularly successful.  The best one can hope for
when comparing with these older measurements is general qualitative
agreement.  

The asymmetry $\beta$ parameter is bound between $-1$ and $2$.  It
attains a minimum close to $-1$ at the same energy as the
corresponding CM, while at energies away from the CM $\beta\to 2$.
The minimum in $\beta$ is narrower than the corresponding Cooper
minimum in the cross section.  For each AMA considered, the presently
calculated $\b$  parameter attains a minimum at an energy between
the locations of the minima in the results of
\citet{Fink1986}, and \citet{Ong1978}.



The Wigner time delays as functions of the ejection angle and the
ejected electron energy are shown in \Fref{Fig6} for Na $3s$, K $4s$,
Rb $5s$ and Cs $6s$ (from left to right). The smooth Coulomb
component, which is quite substantial near the threshold, is
subtracted from these data. The angular dependence of the time delay
can be understood from \Eref{amplitudes} which expresses the spin-up
and spin-down amplitudes $T^\pm$ as functions of the ejection angle.
Both amplitudes are symmetric about an ejection angle of $90^{\circ}$.
The $T^+$ amplitude peaks near $\q=0$ while its counterpart $T^-$
attains its maximum near $\q=90^\circ$. In addition, $T^+$ displays a
prominent CM while $T^-$ remains rather flat. Notably, in the
weakly-relativistic limit, $D_{j=3/2}\approx\sqrt2 D_{j=1/2}$ and
$T^-$ vanishes. So, for ejection angles away from $90^{\circ}$ the
time delay is defined almost solely by $T^+$ and it displays a very
sharp energy and angular dependence. As the ejection angle becomes
close to $90^{\circ}$, the time delay flattens and even changes its
sign.

\section{Summary and outlook}
\label{Summary}

This research has been prompted by a seeming deviation from a common
rule that relates the photo\-ionization cross section and the
attosecond time delay near the Cooper minima (CM) in the noble gas atoms (NGA).
\cite{Ji2024}.  In valence shells of alkali and alkaline-earth metal
atoms,   the phase in the only open non-relativistic $ns\to Ep$ channel
is defined by the Coulomb field of the ion remainder and the target
specific short-range distorted phase. Both contributions remain flat
and are insensitive to the deep CM of the ionization cross
section. Such a departure from the general CM rule established by
\citet{Ji2024} is alerting and begs for a more thorough investigation.

This apparent contradiction is fully resolved when a relativistic
formulation is employed and the two complementary $ns_{1/2}\to
Ep_{1/2}$ and $Ep_{3/2}$ ionization channels are at play near the
CM. The complex ionization amplitudes in both these channels
experience the phase jumps which match accurately the respective CM
and can be replicated by the logarithmic Hilbert transform of the
corresponding partial ionization cross section. So the general rule
formulated by \citet{Ji2024} is fully restored.

We support this conceptual result by numerical simulations across the
whole sequence of alkaline-earth metal atoms (AEMA) within the
relativistic random phase approximation (RRPA)
\cite{PhysRevA.20.964}. The analogous sequence of alkali metal atoms
(AMA) is treated with the model potential method (MPM)
\cite{Singor2021,Singor2022}. The case of Na $3s$ is analyzed within
both the RRPA and the MPM and thus serves as a bridge between the two
techniques.
The RRPA multi-channel model is fully {\em ab initio} and takes into
account the inter-shell correlation that entangles the $ns_{1/2}\to
Ep_{1/2}$ and $Ep_{3/2}$ ionization channels. As the result, the
phases in these channels display a smooth variation near their
respective CM's similarly to the analogous amplitudes in NGA
\cite{Decleva2020}. However, unlike in NGA, the phase variation in the
two relativistic channels of AEMA occurs in the opposite directions
and the net variation vanishes in the non-relativistic limit when both
of these channels become degenerate.
We ventured to adapt RRPA to the open shell AMA and tested this
adaptation on the Na $3s$ atom with reasonable success. More
consistently, AMA should be treated with the MPM technique and we
applied it across the whole AMA sequence.  Even though the MPM is not
fully {\em ab initio} and employs model potentials, it agrees rather
closely with more advanced theoretical methods when the
photo\-ionization cross sections and the angular asymmetry
$\b$-parameters are compared. This comparison enthuses us with some
confidence and lets us proceed with the phase and time delay
evaluation which has not been reported for the AMA in the
literature. Because the MPM is essentially a single-channel model, it
does not entangle the $Ep_{1/2}$ and $Ep_{3/2}$ channels. Hence their
respective phases do not show a smooth variation typical for a
multi-channel RRPA. 

By employing both the RRPA and MPM techniques, we demonstrate the
systematics of the angular-dependent time delay in AEMA and
AMA. Surprisingly, this systematics is rather similar in both the
atomic series with the third row members, K $4s$ and Ca $4s$,
displaying the sharpest variation of the time delay. This is observed
regardless of the open-shell structure of AMA and the closed-shell
configuration of AEMA.  The time delay in Na $3s$ was analyzed with
both RRPA and MPM techniques the two sets of results are qualitatively
similar. This similarity was also found with unpublished results
\cite{Gorczyca2025}.

Our investigation invites experimental verification of the strong
angular and energy-dependent features of the atto\-second time delay
in the $ns$ valence shells of metal atoms. This time delay runs in
thousands of atto\-seconds and  should be termed more appropriately
the femto\-second time delay. The time-resolved photo\-ionization
studies near CM with a small underlying cross section are
challenging. Nevertheless, considerable progress in this area has
been achieved as can be viewed from the latest report
\cite{li2025photoionization}. So we hope that our theoretical
investigation  will prompt further experimental explorations.

\paragraph*{Acknowledgment:}  

The authors acknowledge Thomas Gorczyca, Chii Dong Lin and Kwok-Tsang
Cheng for their useful input and stimulating discussions. We thank
Thomas Gorczyca and Steve Manson for sharing their unpublished results
on the time delay in sodium \cite{Gorczyca2025}.

This work was supported by resources provided by the Australian
Research Council, the Pawsey Supercomputing Centre, the National
Computing Infrastructure of Australia, and the United States Air Force
Office of Scientific Research, Grant No. FA2386-19-1-4044.

\clearpage


\end{document}